# Carbon Nanotube Field-Effect Transistors With Integrated Ohmic Contacts and High-κ Gate Dielectrics

*Ali Javey[1], Jing Guo[2], Damon B. Farmer[3], Qian Wang[1], Dunwei Wang[1], Roy G. Gordon[4], Mark Lundstrom[2], Hongjie Dai[1*]*

[1] Department of Chemistry and Laboratory for Advanced Materials, Stanford University, Stanford, CA 94305, USA

[2] School of Electrical and Computer Engineering, Purdue University, West Lafayette, IN 47907, USA

[3] Division of Engineering and Applied Sciences, Harvard University, Cambridge, MA 02138, USA

[4] Department of Chemistry and Biochemistry, Harvard University, Cambridge, MA 02138, USA

**Abstract**

High performance enhancement mode semiconducting carbon nanotube field-effect transistors (CNTFETs) are obtained by combining ohmic metal-tube contacts, high dielectric constant $HfO_2$ films as gate insulators, and electrostatically doped nanotube segments as source/drain electrodes. The combination of these elements affords high ON currents, subthreshold swings of ~ 70-80 mV/decade, and allows for low OFF currents and suppressed ambipolar conduction. The doped source and drain approach resembles that of MOSFETs and can impart excellent OFF states to nanotube FETs under aggressive vertical scaling. This presents an important advantage over devices with metal source/drain, or devices commonly referred to as Schottky barrier FETs.

* Email: hdai@stanford.edu

In recent years, intensive research on single-walled carbon nanotube (SWNT) based field-effect transistors (FETs)[1-7] has revealed the excellent properties of these novel materials including ballistics transport[2] and high chemical stability and robustness.[1] Nevertheless, it remains an open question as to what the ultimate nanotube FETs may be in structure and performance, and how to achieve the optimum ON-current and conductance, high ON/OFF current ratios, steep switching and highly scaled gate dielectrics and channels. It has been shown recently that with high work function Pd contacts, one can obtain zero or slightly negative Schottky barriers (SBs) to the valence bands of semiconducting tubes (for diameters $d$>~2 nm).[2] This can improve the ON-currents and afford low drain bias conductance near $4e^2/h$. Steep switching between ON and OFF states for nanotube transistors can be achieved by integration of thin high κ gate dielectrics, which produces subthreshold swings close to the theoretical limit of S ~ $(k_BT/e)\ln(10)$ = 60 mV at room temperature.[1]

Here, we report p-channel nanotube FETs comprised of ohmic Pd-tube contacts and high quality thin $HfO_2$ gate insulator films. The objective is to advance nanotube transistors through the integration of optimum contacts and gate dielectrics, a task that has not yet been undertaken previously. The structure of our nanotube FETs is shown in Fig. 1b. Its operation involves bulk switching of the segment of a nanotube underneath an Al top-gate/$HfO_2$ gate stack, while outside the top-gate region the two segments of the tube are electrostatically 'doped' by a back-gate and acting effectively as source and drain (S/D) electrodes. Such nanotube device structure (named 'DopedSD-FETs' here) has been made previously[1,8] mainly for the demonstration of bulk nanotube switching that differs from SB modulation in nanotube FETs with metal as S/D (Fig.1a, denoted as MetalSD-FETs).  However, integration of ohmic contacts for ON-state optimization and detailed characteristics of the OFF-states of DopedSD-FETs have not been addressed previously.

The fabrication of our DopedSD-FETs was similar to that described in ref. 1, involving first the formation of MetalSD-FETs (Fig. 1a) on $SiO_2$ ($t_{ox}$=10nm)/$p^+$ Si substrates.  Pd was used in place of Mo to contact nanotubes here.  The Pd MetalSD-FETs (Si as back-gate) were characterized by electrical transport measurements before subjected to atomic layer deposition (ALD) of a ~8 nm thick $HfO_2$ ($\kappa$~20) film using an alkylamide precursors at 150 ˚C.[9]  Top-gate electrodes were then patterned to afford DopedSD-FETs (Fig. 1b).  Note that in ref. 1, ALD of $ZrO_2$ at 300 ˚C using $ZrCl_4$ as precursor was employed for dielectric deposition.  Compared to the $ZrCl_4$/300 ˚C ALD approach, the alkylamide/150 ˚C approach is advantageous in two respects.  First, the chloride precursor tends to cause irreversible (by e.g., annealing) unintentional p-doping of the nanotubes, resulting in depletion mode FETs.  The alkylamide ALD process does not cause such doping effect, especially after an annealing step (at 180˚C for 2 h) following the deposition.  Second, ALD at 300 ˚C tends to degrade the Pd-SWNT contacts and cause a significant increase in contact resistance.  Such degradation is avoided by ALD at 150 ˚C.

The simultaneous integration of high-$\kappa$ gate dielectrics and high quality Pd-tube contacts affords the highest performance DopedSD-FETs thus far (with back-gate set at a constant bias of $V_{GS\_BACK}$ ~ -2 V).  Fig. 2 shows a representative device (tube diameter $d$ ~ 2.3±0.2 nm) exhibiting a transconductance $g_m$=$dI_{DS}/dV_{DS}|_{V_{DS}}$ ~ 20 µS (corresponding to 5,000 S/m, normalized[7] by 2d; ON-current $I_{ON\_sat} \sim 15\mu A$ (~3750 µA/µm), and a linear ON-conductance of $G_{ON} \sim 0.1 \times 4e^2/h$. A rough estimation shows that the observed $g_m$ and $I_{ON\text{-}sat}$ are higher than the state-of-the-art Si p-MOSFET[10] by a factor of ~ 5 at a similar gate overdrive, and better than previous DopedSD-FETs (with Mo electrodes)[1] by a factor of ~3 due to the improved Pd-tube contacts. The subthreshold swing of the device is S~80mV/decade. The minimum current ($I_{MIN}$) in $I_{DS}$-$V_{GS}$ is relatively bias independent (for $V_{DS}$=0.1 to 0.3 V) and $I_{ON}/I_{MIN} > 10^4$. Beyond the $I_{MIN}$ point, ambipolar n-channel conduction is observed for this $d$~2.3 ± 0.2 nm SWNT device with $I_{ON}/I_{n\text{-channel}}$ close to $10^3$.

We observe comparable p-channel ON-states for our Pd MetalSD-FETs and DopedSD-FETs (i.e., same devices before and after ALD and top-gate formation respectively) with similar $I_{ON}$~15-20 µA and $G_{ON}$~0.1 × $4e^2/h$. This suggests that high-κ deposition does not cause degradation of the ON-states.[1] Since relatively long tubes with L ~ 2 µm are used in this work, the channel transmission is $L_{mfp}/(L+L_{mfp})$~0.1 (non-ballistic channel) where $L_{mfp}$ ~ 300 nm is the mean free path for scattering in nanotubes at low drain biases.[2] We note that in the future, channel length scaling should include both the top-gated tube section and the S/D segments to the ballistic regime (L<~10 nm)[12] to minimize the parasitic resistance. Novel lithography approaches and self-aligned process will be necessary to achieve this goal.

In the subthreshold region, S ~ 70-80 mV/decade for our DopedSD-FETs and S ~ 130 mV/decade for our MetalSD-FETs. The difference appears to be due to the more efficient electrostatic gating for the high-κ/top-gate stack. The top and back gate capacitances are $C_{top}$ ~ 2.9pF/cm and $C_{back}$ ~ 0.38pF/cm respectively as extracted by numerically solving[13] the two-dimensional Poisson equation for a slice in the direction normal to the nanotube. The top-gate capacitance of $C_{top} \sim 2.9$ pF/cm is in fact near the quantum capacitance of $C_Q \sim 4$ pF/cm for SWNTs.[1,3]

It is interesting to compare the minimum currents $I_{MIN}$ and n-channel leakage currents for various types of nanotube FET geometries. First, we note that the back-gated MetalSD-FETs (for tubes with d ≥ 2nm, $t_{ox} \sim 10$ nm) exhibit strong ambipolar conductance with high n-channel leakage currents even under a low bias of $V_{DS}$=10 mV (Fig. 2a inset). This differs from our previous MetalSD-FETs with thicker $SiO_2$ ($t_{ox}$=67 nm) that display negligible n-channel leakage currents.[2] The high n-channel currents for the $t_{ox}$=10 nm case is due to tunneling currents through the thin SB (since width of SB ~ dielectric thickness $t_{ox}$) to the conduction band (CB) of the nanotube.[13-17] For thick gate oxides, the minimum current is determined by thermal activation over the full bandgap of the tube,

$$I_{MIN} \propto \exp(-E_g/k_B T), \qquad (1)$$

which can afford $I_{ON}/I_{MIN} \sim 10^6$ even for d>3 nm ($E_g$ <0.4 eV) tubes under high $V_{DS}$.[2] While the ambipolar conduction and minimal leakage current can be suppressed by producing highly asymmetric Schottky barrier heights for electrons (SB height ~ $E_g$) and holes (SB height ~ 0) when the gate oxide is thick, Fig. 3a clearly shows that the situation is different for thin gate oxide MetalSD-FETs due to high tunneling currents. With aggressively scaled $t_{ox}$ and highly transparent SBs, the minimum current is governed by electron and hole thermal activation barriers (see Fig. 3b) of $\Delta_n = \Delta_p = (E_g - e|V_{ds}|)/2$,[13]

$$I_{MIN} \propto \exp[-(E_g - e|V_{ds}|)/2k_B T)] \qquad (2)$$

For our $t_{ox}$~10 nm MetalSD-FETs (though not yet scaled to $t_{ox}$ ~ 2 nm), we indeed observe the trend of higher $I_{MIN}$ for higher $V_{DS}$ (Fig.3a). Note that it has been pointed out recently[13,15] that as a result of Eq. 2, ultra-scaled MetalSD-FETs will exhibit unacceptable OFF state leakage under useful operating voltages (e.g., $V_{DS}$ ~ 0.6 V) even for devices with small diameter tubes (d~1 nm, $E_g$ ~ 0.8 eV).

Our DopedSD-FETs exhibit low $I_{MIN}$ and much suppressed ambipolar conduction (relative to the MetalSD-FETs, Fig. 2a inset) for $V_{DS}$ = 0.1-0.3 V (Fig.2a). No significant bias dependence for $I_{MIN}$ is observed, at least for $d$<~2.3 nm, and high $I_{ON}/I_{MIN}$ ~ $10^5$ can be readily obtained at $V_{DS}$ = 0.3 V. These characteristics are drastically improved over the MetalSD-FETs owing to the design of using nanotube segments as S/D, as predicted theoretically.[13] The minimum leakage current is predicted to be determined by a hole activation barrier of $\Delta_p$ ~ $(E_g - E_d)$,

$$I_{MIN} \propto \exp[-(E_g-E_d)/k_BT] \qquad (3)$$

where $E_d$ is the energy spacing from the valance band edge to the Fermi level in the p doped S/D segments (Fig 2b).[13] In our experiments, $E_d$ is set by back-gate electrostatic doping under $V_{GS\_BACK}$ ~ -2 V, which corresponds to $E_d$ ~ 0.2 eV (back gate efficiency for $t_{ox}$=10 nm SiO$_2$ is ~ 0.1). Eq. 3 also predicts the insensitivity of $I_{MIN}$ to $V_{DS}$, as observed experimentally. The fundamental difference for the OFF state characteristics between the Metal- and DopedSD-FETs is that the latter employs doped semiconductors as S/D, much like a conventional MOSFET. Since there are no states within the bandgap of the S/D electrodes, $I_{MIN}$ will be determined by activation over ~ $E_g$ as opposed to $E_g/2$ in the MetalSD-FETs. The n-channel leakage current in the DopedSD-FETs is due to band-to-band tunneling (Fig. 2b, bottom drawing), which is low since high gate-voltages are required to obtain $I_{n\text{-channel}}$ comparable to $I_{ON}$. The results here clearly demonstrate that DopedSD-FETs are much more vertically scalable than MetalSD-FETs, and can afford low OFF state current even for relatively large diameter (~ 2 nm) tubes and high biases.

We have also characterized DopedSD-FETs for SWNTs with diameter ~ 1.5 nm. The measured transfer characteristics of a $d \sim 1.5$ nm ($E_g \sim 0.6$ eV) DopedSD-FETs is shown in Fig. 4. Due to the presence of small but finite SBs at the Pd-tube contacts for d< 2 nm tubes[2] and thus higher parasitic resistance, relatively low $G_{ON} \sim 0.02 \times 4e^2/h$ is measured for this device. Nevertheless, the device exhibits $S \sim 80$ mV/decade, as a result of near MOSFET operation (instead of SB modulation). We observe high $I_{ON}/I_{OFF} \sim 10^5$ and no ambipolar conduction, suggesting excellent OFF states for DopedSD-FETs with small diameter nanotubes (but at the expense of lower ON-states).

In summary, enhancement mode nanotube FETs with high quality contacts, high-κ dielectric $HfO_2$ films and electrostatically induced nanotube source/drain regions are demonstrated. Future tasks will include chemical doping of the nanotube S/D segments to replace back-gate electrostatic doping. Contacts with nearly zero-SBs to small diameter SWNTs should be developed to optimize $I_{ON}/I_{OFF}$ during vertically scaling. Strategies for channel length scaling for DopedSD-FETs should also be devised.

This work was supported by the MARCO MSD Focus Center, DARPA Moletronics, NSF Network for Computational Nanotechnology and a SRC Peter Verhofstadt Graduate Fellowship (A. J.).

**Figure Captions:**

**Figure 1**. a) Schematic device drawings for a nanotube FET with metal as S/D (MetalSD-FETs, thickness of $SiO_2$ = 10 nm). b) A nanotube FET with back-gate electrostatically doped nanotube segments as S/D (DopedSD-FET). The thickness of the top Al gate is ~ 20 nm. c) Scanning electron microscopy (SEM) image of a device depicted in b). For all of our devices here, the total tube length between metal electrodes was ~ 2 µm, the top-gated section length was ~ 0.5 µm. Misalignment caused differences in the lengths of the S/D tube segments.

**Figure 2**. Electrical properties of a DopedSD-FET (tube diameter ~ 2.3 nm). a) Transfer characteristics at different $V_{DS}$ (dashed curve $V_{DS}$=–0.1V; dotted: $V_{DS}$=-0.2 V; solid $V_{DS}$=–0.3V). Inset: the same tube versus back-gate under $V_{DS}$=10 mV in MetalSD-FET geometry prior to high-κ deposition. b) Band diagrams corresponding to $I_{ON}$ (top), $I_{MIN}$ (center) and $I_{n\text{-channel}}$ (bottom) regions in (a). The shaded region corresponds to the top-gated nanotube section. c) Output characteristics of the top-gated device in a).

**Figure 3**. a) $I_{DS}$-$V_{GS}$ curves for a MetalSD-FETs (d~2.3 nm, back-gated, $SiO_2$ thickness 10 nm) under various $V_{DS}$ (solid line $V_{DS}$=0.3 V; dotted: 0.2 V; Dashed: 0.1V). b) Band diagrams for the device under gate biases corresponding to the n-channel (top diagram) and minimal (bottom diagram) leakage currents respectively.

**Figure 4.** Transfer characteristic of a DopedSD-FET with a *d*~1.5 nm SWNT.

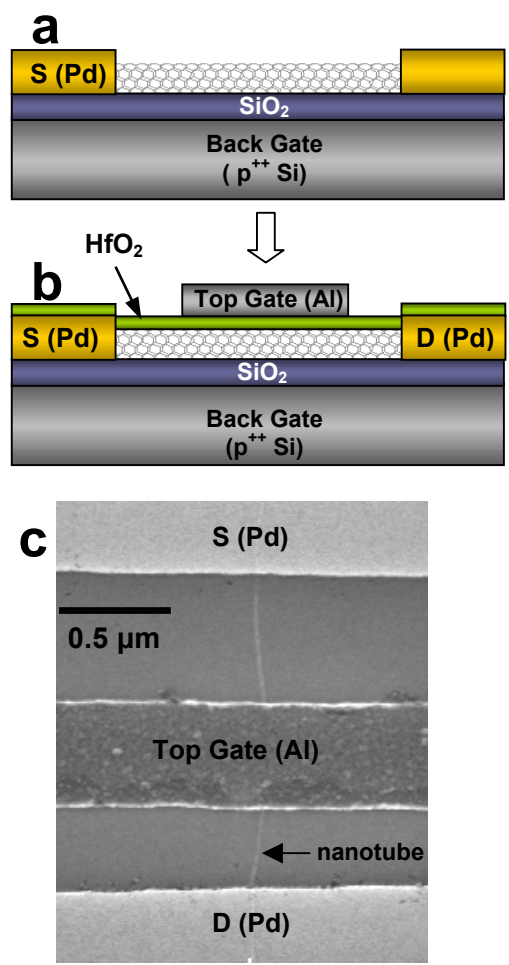

Figure 1

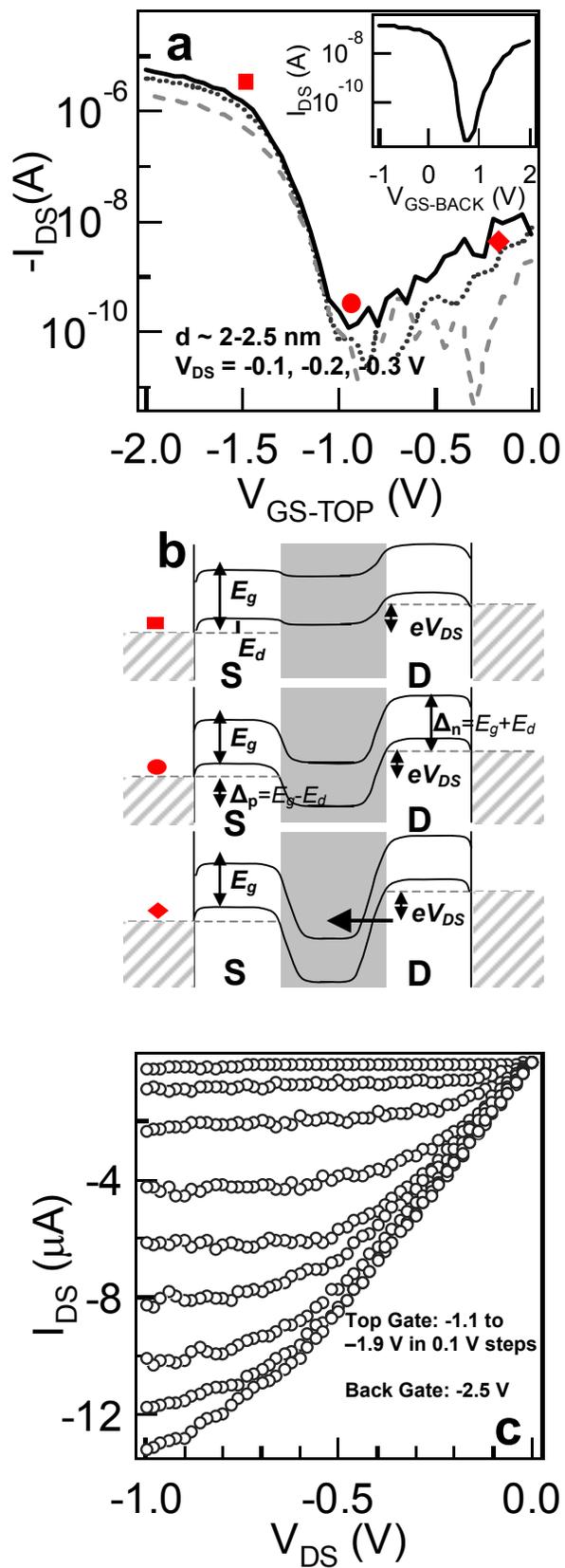

**Figure 2**

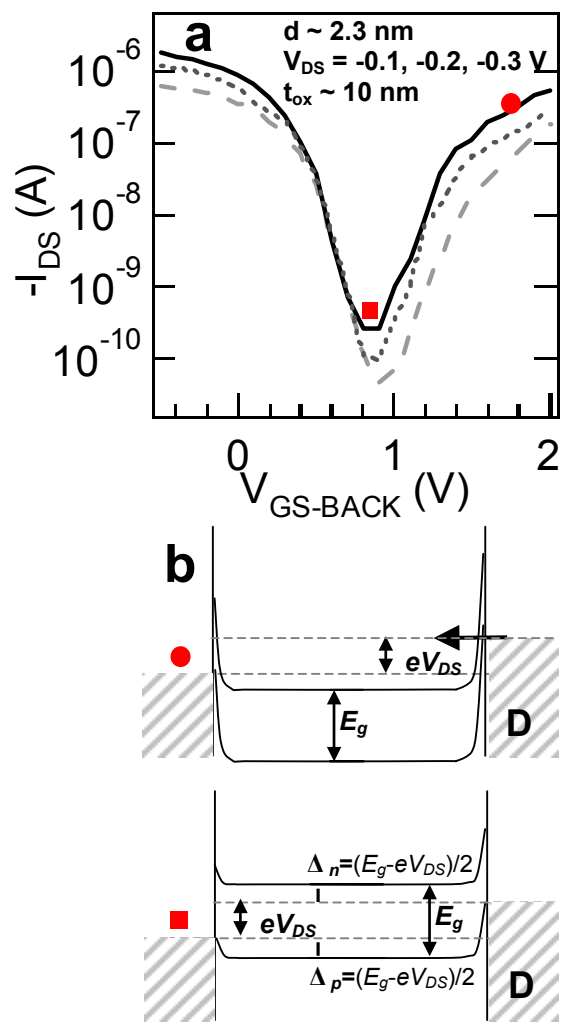

**Figure 3**

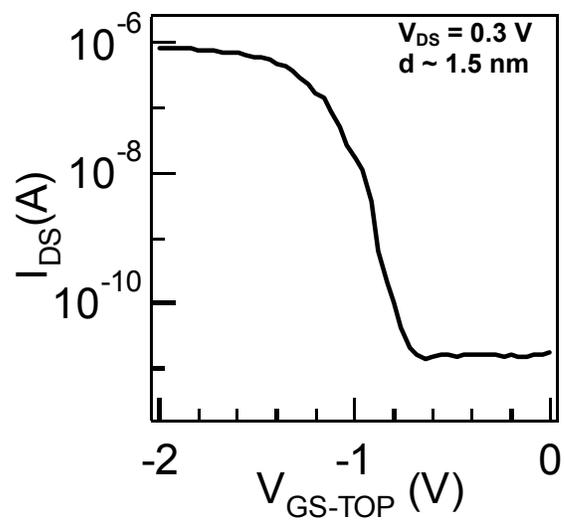

**Figure 4**